\pdfoutput=1 
\documentclass[12pt]{article}
\usepackage{amsmath,amsfonts,amssymb,url,color,epsfig,graphics}
\usepackage[latin1]{inputenc}
\newcommand{\R}{{\mathbb{R}}}

\newcommand{\C}{{\mathbb{C}}}
\newcommand{\Z}{{\mathbb{Z}}}

\def\ha{\frac{1}{2}}

\def\pa{\partial}
\def\ra{\rightarrow}

\def\nghbd{neighbourhood~}

\newtheorem{defi}{Definition}[section]

\newtheorem{lemm}{Lemma}[section]
\newtheorem{prop}{Proposition}[section]
\newtheorem{rem}{Remark}[section]

\newtheorem{theo}{Theorem}[section]

\newenvironment{demo}{\noindent {\it Proof.--}
      \begin{quotation}\noindent}{\end{quotation}\hfill$\square $}

\begin{document}
%\large

\title{Spectral theory of pseudo-differential operators   of degree $0$ \\ 
and application to  forced linear waves}

\author{Yves  Colin de Verdi\`ere\footnote{Universit\'e Grenoble Alpes,
Institut Fourier,
 Unit{\'e} mixte
 de recherche CNRS-UGA 5582,
 BP 74, 38402-Saint Martin d'H\`eres Cedex (France);
{\color{blue} {\tt yves.colin-de-verdiere@univ-grenoble-alpes.fr}}}}

\maketitle

\section*{Introduction}

This paper contains new developpements of some ideas already introduced in our paper \cite{CSR-18} concerning the spectral theory of 
self-adjoint pseudo-differential operators of degree $0$ on closed manifolds.
The main motivation comes from the study of forced internal or inertial  waves in physics, see 
\cite{BFM-13,Br-16,GDDSV-06,MBSL-97,ML-95,Og-05,RV-10,RV-18,Pill-18}
and many other works.
In what follows, 
 $H$ is a classical self-adjoint scalar pseudo-differential operator of degree $0$ on a compact
 manifold $M$ of dimension $n$ without boundary,
 $f$ is a smooth function and the spectral parameter  $\omega $ is a real number. 
The main object to study is the following  linear forced wave equation:
\begin{equation}\label{equ:forced}
\frac{1}{i}\frac{du}{dt}+ Hu =fe^{-i\omega t},~u(0)=0~. \end{equation}
We are interested in the behaviour of 
$u(t)$ as $t\ra +\infty $.  Thanks to the spectral theorem, we can relate this behaviour to the spectral theory of $H$
and hence to the Hamiltonian dynamics of the principal symbol $h:T^\star M \setminus 0\ra \R $ which is a smooth function homogeneous
of degree $0$. 
The main tools that we use are already classical: they are, on one hand, the general theory of pseudo-differential operators, culminating
in the works of Lars H\"ormander, Hans Duistermaat, Alan Weinstein  and many others, in the beginning of the seventies,
see \cite{Du-11,Fo-89,We-71,We-75,DZ-17}, and,  on the other hand, the  theory
initiated by Eric Mourre in the beginning of the eighties in order to get a flexible way
to have  a limit absorption principle, see \cite{Mo-81,Mo-83,JMP-84,Ge-08,Ca-05}.

What is the content, beyond that of \cite{CSR-18}?
The main result is Theorem \ref{theo:main} where we extend the result of \cite{CSR-18} to the generic Morse-Smale case stil in 
dimension $2$. The other new contribution is a precise description in arbitrary dimension 
 of the dynamical assumptions allowing to apply Mourre theory thanks to the  G\r{a}rding inequality
(see Section \ref{sec:escape}) by constructing a {\it global escape function.}

After recalling general facts on the Hamiltonian dynamics of a homogeneous Hamiltonian $h$ of degree $0$
in Section  \ref{sec:ham-class} and on the spectral theory of $H$  in Section  \ref{sec:ham-sp}, we give,
in  Section \ref{sec:escape},  a necessary and sufficient
condition on the dynamics at infinity, which insures the existence of an escape function that  will be the key input in order to apply
 Mourre's theory thanks to the  G\r{a}rding inequality. 
In Section \ref{sec:mourre}, we  recall
general facts that  we got in \cite{CSR-18}  for the forced wave equation from Mourre's theory.
In Section \ref{sec:DZ}, 
we use radial propagation estimates (see \cite{DZ-17,DZ-18}), going back to works of Melrose and Vasy, in order
to locate the wavefrontset of the Schwartz distribution $u_\infty $ which is the limit (modulo bounded functions in $L^2$)
 of $u(t)$ as 
$t\ra +\infty $. 

We consider then,  in Section \ref{sec:2D}, the case where $M$ is a  surface ($n=2$),
 extending our results of \cite{CSR-18} to the generic case where the foliation is Morse-Smale and can 
have  singular points  (foci, nodes or saddles). 
Finally, we consider, in Section \ref{sec:3D},  the case where $M$ is a 3D manifold with a free  $S^1-$action
 leaving $H$ invariant which is important for  applications to physics. 
We end the paper with a short review of related problems in Section \ref{sec:pb} and two Appendices. 

{\it Aknowledgments: I would like first to  thank Laure Saint-Raymond for proposing to me to work
 with her on this exciting  problem and for sharing many
ideas.  Other people were helping me  at different steps  of this work:
Thierry Dauxois, Semyon Dyatlov, Fr\'ed\'eric Faure, Vladimir Georgescu,  Etienne Ghys, Johannes Sj\"ostrand, Michal Wrochna,
Maciej Zworski,  I thank all of them very warmly.

Many thanks also to the referee whose detailed report including many pertinent remarks  allowed me to improve the paper.}

\section{Hamiltonian of degree 0: classical dynamics}\label{sec:ham-class}
In what follows, we fix the following notations:
$M$ is a smooth connected compact manifold of dimension $n\geq 2$ without boundary, $q$ is the generic point
 of $M$ and $|dq|$ a smooth density
on $M$. The Hamiltonian $h$ is a smooth positively homogeneous function $h:T^\star M\setminus 0 \rightarrow \R$.
 We denote by $(q,p)$
some local canonical coordinates on $T^\star M$ and by extension a generic point of $T^\star M$.
The Hamiltonian vector field of $h$ is denoted by 
${\cal X}_h$ and we fix the ``symplectic''  conventions so that
\[ {\cal X}_h =\frac{\pa h}{\pa p}\pa _q-\frac{\pa h}{\pa q}\pa _p ,~  {\cal X}_h f=\{ h,f \} \]
and denote by 
$\Phi_t $ the flow of ${\cal X}_h$.
Because of the homogeneity of $h$, we have
$pdq  ({\cal X}_h)=0$ and ${\cal X}_h$ is homogeneous of degree $-1$.
Let us fix $\omega \in \R $ and define the energy shell 
$\Sigma _\omega := h^{-1}(\omega )$. We will assume in what follows that $\omega $ is {\bf not a critical value of} $h$ and hence $\Sigma _\omega $
is a smooth conic hypersurface in $T^\star M\setminus 0$. We need to introduce $Z_\omega:=\Sigma _\omega /\R^+$ which 
 is a smooth closed manifold of dimension $2n-2$ and  will  be seen 
as the boundary at infinity of $\Sigma _\omega $.
 The vector field ${\cal X}_h $ defines by projection a conformal class
of vector fields on $Z_\omega$, which we will call  an (oriented) foliation and  denote by ${\cal F}$. 
This foliation can admit  singular points corresponding to the lines $\R^+. (q,p)$ where 
${\cal X}_h $ is parallel to the cone direction $p\pa _p$. 
Note that we can and will often  reduce ourselves to the case  $\omega =0$ by looking at the Hamiltonian $h-\omega$.

\section{Hamiltonian of degree 0: spectral theory}\label{sec:ham-sp}
Let us choose a self-adjoint pseudo-differential operator $H$ of degree $0$ acting on
$L^2(M,|dq|)$ and of principal symbol $h$. Note that $H$ is a bounded operator. 
In what follows, all pseudo-differential operators are ``classical'', it  means that the symbols do have
full expansions in homogeneous functions with integer degrees. 
We are mainly interested by the spectral theory of $H$.
As a warm up, we have the 
\begin{theo} The essential spectrum of $H$ is the interval $J:=[h_-,h_+]$
with $h_-:=\min h,~h_+:=\max h $.
\end{theo}
{\it Proof.--}
If $\omega \in \C \setminus  J$, $H-\omega $ is elliptic and hence admits an inverse
$R(\omega )$  modulo compact operators which   can be  chosen  holomorphic in $\omega $
by taking $R(\omega ):= {\rm Op}(h-\omega )^{-1}$ where ${\rm Op}$ is a fixed quantization on $M$: 
\[ R(\omega )(H-\omega )={\rm Id }+ K(\omega ) \]
with $K$ compact  and holomorphic in $\omega $. On the other hand, $H$ being bounded,  $(H-\omega )$ is invertible for 
large values of $\omega$.  It follows from the Fredholm analytic Theorem that
the operator $H-\omega $ is invertible outside a discrete set where the kernels are finite dimensional. 

On the other hand, if $\omega \in J$, with $h(q_0,p_0)=\omega $ and $\epsilon >0$ is fixed,
choose a small neighbourhood $U $ of $q_0$ so that, if $q\in U$, $|h(q,p_0)-\omega |\leq \epsilon $.
Pick then  $\phi \in C_o^\infty (U) $ with $\int _M |\phi|^2 (q) |dq|=1$.
Let us check that, for $t$ large enough,
\begin{equation} \label{equ:quasi} \| (H-\omega )\left(\phi e^{itq p_0 }\right) \|_{L^2 (M)}\leq 2\epsilon \end{equation}
It follows from the general properties of the principal symbols that
\[ H\left(\phi (q) e^{itq p_0 }\right)=h(q,p_0)\phi  (q) e^{itq p_0 }+ O\left( \frac{1}{t} \right) \]
Take $t$ so that the $L^2$ norm of the remainder is smaller than $\epsilon $.
We get inequality (\ref{equ:quasi}) by applying the triangular inequality. 
Hence
\[ \|   (H-\omega )\left(\phi e^{itq p_0 } \right)\|_{L^2 (M)}\leq 2\epsilon \| \phi e^{itq p_0} \|_{L^2 (M)} \]
which proves that $\sigma (H) \cap [\omega -2\epsilon, \omega +2 \epsilon] \ne \emptyset $.

$\square$

\section{Escape functions}\label{sec:escape}

The key object of this paper   is an   escape function for $h$ on the energy shell $\Sigma _0$: 
\begin{defi} \label{defi:escape}
 A smooth  function $k:\Sigma _0 \ra \R $, positively homogeneous of degree $1$,
is called an {\rm escape function} if there exists $\delta >0$ so that the Poisson bracket
$\{ h, k\}={\cal X}_h k$ is larger than $\delta$ on $\Sigma _0$.  
\end{defi}
A key observation is: 
\begin{rem} If we extend $k$ to $T^\star M \setminus 0$ as a smooth function $\tilde{k}$ homogeneous of degree $1$, then
$\tilde{k}$ restricted to $\Sigma _\omega $ is stil an escape function on $\Sigma _\omega $ for $\omega $ small enough.
\end{rem}

We first give   a general dynamical assumption on the oriented foliation ${\cal F}$ which turns  out to be equivalent
to the existence of a global  {\it escape function}.
We need some definitions, using the definitions of Appendix \ref{sec:dyn}:
\begin{defi}
We will say that the oriented 1D foliation ${\cal F}$ of the manifold $Z_0$  admits a  {\rm simple} structure
$(K_+,K_-)$ if 
 $Z_0=K_+ \cup K_- \cup \Omega $ as a disjoint union where:
\begin{itemize}
\item $K_+$ is  an  attractor  of the  oriented foliation ${\cal F}$, the {\it sink}
\item $K_-$ is a  repellor    of the  oriented foliation ${\cal F}$, the {\it source}
\item  All  leaves of points in $\Omega $  converge to $K_+$ at ``$+\infty $'' and to $K_-$ at
``$-\infty $''; in particular,  the basin of $K_+$ is $\Omega \cup K_+$ and the basin of $K_-$ for the reversed orientation 
of  ${\cal F}$ is $\Omega \cup K_-$.
\end{itemize}
\end{defi}
and 
\begin{defi}
We say that a compact invariant set $K_+$ is weakly hyperbolic, denoted {\bf (WH)},
if there exists, in some neighbourhood of $K_+  $, a vector field $W$ generating ${\cal F}$ and a smooth density
$d\mu $ so that ${\rm div}_{d\mu} (W) <0 $. Similarly for $K_-$, 
 ${\rm div}_{d \mu} (W) >0 $.
\end{defi}

Our main result in this section is  
\begin{theo} \label{theo:escape} If the foliation ${\cal F}$ has a  simple structure  $(K_+ , K_- )$ with  
 $K_+ $ and  $K_-$ satisfying {\bf (WH)}, then there exists an escape function.

The converse is true: the existence of an escape function implies that the foliation ${\cal F}$ has a  simple
structure  $(K_+ , K_- )$   so that
 $K_+ $ and of $K_-$ satisfy {\bf (WH)}. This simple structure is uniquely determined by  ${\cal F}$.
\end{theo}

\subsection{Dynamical assumptions implying   weak hyperbolicity}\label{sec:wh}
Let us choose a vector field $W$ generating ${\cal F}$, whose flow is denoted by $\phi_t,~t\in \R$,  and  equip 
  $Z_0$  with a smooth density $d\mu $.

Let us describe properties of closed invariant sets of ${\cal F}$ from which we can deduce  {\bf (WH)}:
\begin{enumerate}
\item
If some component of $K_+ $ is an  isolated point $a$, the assumption {\bf (WH)} says that the trace  of the 
 linearized  vector field of $W$  at the point $a$  is negative.
This is independent of the choice of $W$. 
The case where  the singular point is hyperbolic is studied in the work of Guillemin and Schaeffer \cite{GS-77}. They show that,
in the generic situation, there exists a pseudo-differential normal form for such points.
Independently,  the classical part of this  normal form is also described in dimension  2
in the works of Davydov and co-authors \cite{Da-85,Ar-88,DIIS-03}. 
\item
If some component of $K_+ $  is a closed curve $\gamma $, the assumption {\bf (WH)} says   that
  the modulus of the determinant of  the linearized Poincar\'e map  is  $<1$.   In dimension $n=2$, this is equivalent to
our assumption $(M2)$ in \cite{CSR-18}. 
\item They are more complicated attractors which satisfy {\bf (WH)}. The Lorenz attractor is one of them: the vector field
generating it has negative divergence. 

\end{enumerate}

%%%%%%%%%%%%%%%%%%%%%%%%%%%%%%%%%%%%%%%%%%%%%%%%%

\subsection{Construction of an  escape function}\label{sec:build-escape}

We construct an escape function assuming that  ${\cal F}$ has a simple structure with $K_\pm $ satisfying  {\bf (WH)}.

\subsubsection{Escape function near  $\Gamma _+$}\label{sec:wh-escape}

Let $\Gamma _\pm $ be the sub-cones of $\Sigma _0$ generated by the sets $K_\pm $.
We will  construct in this section an escape function $k_+$ in some conic neighbourhood $U +$ of   $\Gamma_+$. 
A similar construction can be done on the basin of $\Gamma _-$.

Let us first construct  ``polar coordinates'' $(\rho, \theta )$ on $\Sigma_0 $ where $\rho \in \R^+\setminus 0$,   $\theta \in Z_0$ and
the dilations on $\Sigma_0 $ act by $\lambda .(\rho,\theta )=(\lambda \rho, \theta )$: 
\begin{lemm}
If $W$ is a given vector field on $Z_0$
generating ${\cal F}$, 
 there exist polar coordinates $(\rho, \theta )\in (\R^+\setminus 0) \times Z_0 $
on $\Sigma _0 $ so that \[ {\cal X}_h = a(\theta )\pa _\rho + \frac{1}{\rho}W~. \]
\end{lemm}
{\it Proof.--}
  We start with arbitrary polar coordinates $(\rho_1,\theta )$:
for example identify $Z_0$ with the co-sphere bundle $S^\star _1$ for some Riemannian metric on $M$ and define $\rho_1 (q,p)$ so
that $(q, p/\rho_1(q,p))\in S^\star _1$.
 We  get, using the homogeneity of $ {\cal X}_h$ and
the fact that $W$ span ${\cal F}$, 
\[ {\cal X}_h = a_1(\theta )\pa _{\rho_1} + \frac{1}{\rho}W~\]
with $\rho =A(\theta )\rho_1 $
and hence
$\pa _{\rho_1} =A(\theta )\pa _{\rho}$. $\square$

The Liouville measure $dL_0:=|dq dp/ dh | $ on $\Sigma _0 $, being homogeneous of degree $n$,  w.r. to dilations,  writes 
$dL_0 = \rho ^{n-1}|d\rho | d\mu $ where $d\mu $ is a smooth measure on $Z_0$.

The fact that 
\[ {\rm div}_{dL_0}( {\cal X}_h)=0 \]
rewrites
\begin{equation} \label{equ:equ-div}  (n-1) a + {\rm div}_\mu (W) =0 \end{equation}
The assumption {\bf (WH)} implies that we have a smooth $>0$ function $F$ defined  near    $K_+$ so that
\[ {\rm div}_{F\mu }(W)=\frac{dF(W)}{F} + {\rm div}_\mu (W) \leq -c <0 \]

Then, if $k_+ :=F^{-1/(n-1)} \rho $,
we get
\[ dk_+ ({\cal X}_h )= -\frac{1}{n-1}F^{-1/(n-1)}\left(\frac{dF (W)}{F}-(n-1)a \right)  \]
which is equal to
\[  dk_+ ({\cal X}_h )=-\frac{1}{n-1}F^{-1/(n-1)} {\rm div}_{F\mu }(W) \]
and
we get 
that the  function   $k_+$ is an escape function in some conical neighbourhhood of $\Gamma _+$.     $\square $

We define similarly $k_-:=-F^{-1/(n-1)} \rho $.

Note that $k_+ $ tends to $+\infty $ as $z $ tends to $K_+ $ viewed as a set of points at infinity of $\Sigma _0$. 
We have also $k_+ \sim <p > $ from the definition and the fact that $F$ is positive.  

Similarly, the function $k_- $ defined near $\Gamma _-$ tends to
$-\infty $ as $z $ tends to $K_-$.

\subsubsection{Extension to $\Sigma _0$}\label{ss:escape}

We  choose a positive   function $m$ on $\Sigma _0$  which is smooth, homogeneous of degree $0$ and
equal to  $m_\pm:=\{h, k_\pm   \} $ in  some conical neighbourhoods $U_\pm $ of  $\Gamma _\pm$.
It follows from Item 3 of Proposition \ref{prop:att} that we can choose $U_+ $ so that $\Phi_t(U_+)\subset U_+$ for $t\geq 0$ 
and similarly for $U_-$. 

Let $z$ be in the basin of $\Gamma _+$ and 
 define 
\[ l_+(z)=\lim _{t\ra +\infty } \left( k_+(\Phi_{t}(z))-\int _0^{t} m(\Phi_s (z) ) ds \right)\]
The limit exists because the expression of which we take the limit is independent of $t$ for $t$ large enough. 
Moreover the limit is smooth: if $z$ is given and $\Phi_T(z)\in U_+$ for all $T\geq T_0$, there exists a neighbourhood $V$ of $z$ so that
 $\Phi_{T_0}(V)\subset U_+$ and hence  $\Phi_{T}(V)\subset U_+$ for all $T\geq T_0$. We have then, for $w\in V$,
\[ l_+(w)= k_+(\Phi_{T_0}(w))-\int _0^{T_0} m(\Phi_s (w) ) ds \]
which is clearly smooth.

We define similarly $l_-$. 
The functions   $l_\pm $ are escape functions in   the basins  of $\Gamma _\pm$ and satisfy in the respective basins
$\{h,  l_\pm \}=m  $.

 Let $\Gamma_0$ be  the cone $\Gamma_0:=\{l_+=0\}$ which is smooth and transversal
to ${\cal X}_h$ because $dl_+ ({\cal X}_h )=m >0 $.  
 On  $\Gamma_0 $ we have now the two functions $l_\pm $. The difference $\delta (z)=l_+ (z) -l_- (z) $
is homogeneous of degree $1$ and is constant along the flow lines. 
We will define  $k$ on the Hamiltonian trajectories $t\ra \Phi_t(z)$  starting from $z\in \Gamma_0^1:=\{ g^\star =1 \}\cap \Gamma_0$. 
For further use, we denote by $S$ this hypersurface of $T^\star M$. 
The set $\Gamma_0^1$ is compact and hence the function $|\delta |$ is bounded by some constant $C>0$ on it.
Let us put $m_0:=\min m >0$ and let $\psi :\R \ra \R$ be a smooth function satisfying
\begin{itemize}
\item $\psi (t)=0$ is $t\leq 0$
\item $\psi (t)=1$ if $t \geq 4C/m_0$
\item $|\psi '|\leq m_0/2C $ 
\end{itemize}
We define now for $z \in \Gamma_0^1$, 
\[ k(\phi_t(z))= (1-\psi(t)) l_- (\Phi_t(z)) +\psi(t) l_+ (\Phi_t(z)) \]
The derivative of $k$ with respect to ${\cal X}_h$ is then
equal to $m + \psi' (l_+ -l_-) \geq m_0/2 $. 
 We extend then $k$
by homogeneity.

\subsection{Deriving  the properties of ${\cal F}$  from   the existence of an escape function}
In what follows, we assume only the existence of an escape function $k$. 

Let us give  
a construction  of $\Gamma _\pm $ using only the dynamics of ${\cal X}_h$.
We will  see that these sets are defined independently of the choice of
$k$: $\Gamma _+$ is the set of points $z\in \Sigma _0 $ so that
there exists $t_0<0 $ with $\Phi_t(z) \ra 0 $ as $t\ra t_0^+$, i.e. the trajectory of
${\cal X}_h$ is not complete as time $t\ra -\infty $. 
Similarly for $\Gamma _-$ with $t_1>0$. We define $K_\pm $ so that they generate the cones $\Gamma _\pm$.
Note that $\Gamma _+ \cap \Gamma _- =\emptyset $: ifnot, let   $z\in \Gamma _+\cap \Gamma _-$, then  $\Phi_t (z)$ tends to the zero section
of $T^\star X$  as $t=t_0+0$, because the Hamiltonian flow is complete near the infinity of $T^\star X$.  
 $\Phi_t (z)$ tends also to the zero section as $t=t_1-0$. This is not possible because the escape function tends to $0$
at the zero section  and is monotonic along the orbits. 

Le us recall that we see $K_\pm $ as sets at infinity of the energy shell, namely the bases at infinity of the
cones $\Gamma _\pm$.  
\begin{prop} The picture of the dynamics is as follows:
\begin{itemize}
\item 
if $z\in \Sigma _0 \setminus (\Gamma_+ \cup \Gamma_-)$, $\Phi_t(z)$ is defined for all $t\in\R$,
$\Phi_t (z)\ra K_+$ as $t\ra +\infty $ and
$\Phi_t (z)\ra K_-$ as $t\ra -\infty $
\item 
if $z\in \Gamma_+ $, $\Phi_t(z)$ is defined for all $t>t_0(z)$,
$\Phi_t (z)\ra K_+$ as $t\ra +\infty $ and
$\Phi_t (z)\ra 0$ as $t\ra t_0 (z)$
\item 
if $z\in \Gamma_- $, $\Phi_t(z)$ is defined for all $t<t_0(z)$,
$\Phi_t (z)\ra K_-$ as $t\ra -\infty $ and
$\Phi_t (z)\ra 0$ as $t\ra t_0(z)$.
\end{itemize}
\end{prop}
{\it Proof. --} Let us choose a metric $g$  on $M$ and consider the set 
$C_0:=k^{-1}(0) \cap (g^\star)^{-1}(1)$ where $g^\star $ is the dual metric. The set $C_0$  is a generating set for the cone $C:=k^{-1}(0)$.
If $z\in C_0$, the trajectory $t\ra \Phi_t(z)$ is complete, because  $t \ra k(\Phi_t(z))$ is strictly monotonic
 and
hence does not  tend to the zero section where $k=0$ at $t=\pm \infty $. 
Conversely, every complete trajectory cuts $C_0$ exactly in one point. 
 This way we get a subset  $S$
 of $\Sigma _0 $   generating $\Sigma _0\setminus \left( \Gamma_+ \cup \Gamma_- \right)$:
\[ S:= \{ \Phi_t (z)~|~z\in C_0,~t\in \R \} \]
The orbits sitting in $S$ have no limit points in $S$ because the flow derivative of $k$ is bounded below by some positive number. 
Let us consider the projections on $Z_0$ of $S$, $\Gamma_+$ and $\Gamma _+$, say $\Omega $, $K_+$ and $K_-$.
We have a disjoint union $Z_0= \Omega \cup K_+ \cup K_-$. Each set is invariant by the foliation. 
Let us look at a leaf  $\gamma $  in $\Omega $: $\gamma $ has no limit points in $\Omega $ (because the foliation in $\Omega $
is diffeomorphic to the flow foliation in $C$). The limit points are then in $K_+\cup K_-$. 
We have $\Gamma _+ \subset \{ k>0  \} $ and $ \Gamma _-\subset \{ k<0  \} $. Hence the limit points at $+\infty $ are in $K_+$
and the limit points at $-\infty $ in $K_-$.
The set $K_+ $ is an attractor: it is enough to consider the neighbourhoods $U_N$  of $K_+$ which are the projections
of the sets $\left( \{ k\geq N \}\cap C \right) \cup \Gamma _+$.

 $\square$

Let us show that the existence of an escape function implies that $K_+ $ satisfy {\bf (WH)}: 
we choose  polar coordinates $(\rho, \theta )$ near $\Gamma _+$ with  the  $\rho =k$ and we have, 
from the equations derived in section \ref{sec:wh-escape}, that 
$dk ({\cal X}_h)= a >0$ and hence ${\rm div}_\mu W =-a/n <0 $: all components of $K_+$ satisfy {\bf (WH)}. 
A similar argument works for $K_-$. 

\subsection{Radial sink and sources}\label{sec:source}

Let us recall and introduce some notations: the radial compactification of $T^\star M$ is denoted by 
$\overline{T^\star M}$ and the boundary at infinity which we can identify with the sphere bundle 
is $S^\star M:=T^\star M/\R^+$. The compactification of $\Sigma _0 $ is 
$\overline{\Sigma _0}$ with the boundary at infinity $Z_0=S\Sigma _0 \subset \overline{T^\star M}$.

Let us rephrase the Definition E.52 of \cite{DZ-17}  in our context: 
\begin{defi} Let us introduce the  symbol $r=-kh $, with $k$, an escape function (homogeneous of degree $1$), 
and denote by $\psi_t $ the flow of $r$ extended to the boundary. The compact set 
$K_-\subset  Z_0 $ is a radial source for $r$   if there exists a  neighbourhood $U\subset \overline{T^\star M} $ of $K_-$, 
so that, uniformly for $z\in U$, 
\begin{enumerate}
\item  For $t \leq  0 $,  $|k| (\psi_t (z))\geq C e^{\theta |t|}$ for some $C,~\theta >0$.
 \item   
$\psi_t(z)\ra K_-$ as $t\ra -\infty $.
\end{enumerate}
\end{defi}
We have:
\begin{prop}
If $k$ is  an escape function, 
$K_-$ is a radial  source for $r=-kh $. 
\end{prop}

{\it Proof--.} We have, in the domain where $k<0$, in particular near
$K_-$,  ${\cal X}_r = |k|{\cal X}_h -h {\cal X}_k $. 
The vector field ${\cal X}_r$ is homogeneous of degree $0$ and hence projects onto $S^\star M$. We denote
${Y}_r $ this projection.  Note that  ${Y}_r$ is tangent to $Z_0 $ where it generate the foliation
${\cal F}$.

Let us prove item 1: 
we have ${\cal X}_r (|k|)= |k|{\cal X}_h |k|\leq -\delta |k |$. 
This implies that in a neighbourhood $U_0$ of $K_-$ where $k\leq -1$, we have, for $t\leq 0$, 
 $|k| (\psi_t (z))\geq C e^{\delta  |t|}$. 

Let us prove item 2: 
let us choose $V_0$ a neighbourhood of $K_-$ inside $S^\star M$ as follows:
we choose first a neighbourhood $V_1$ of $K_-$ in $Z_0$, with a smooth boundary,  so that ${\cal Y}_r $
 is outgoing and transversal to the boundary: take $V_1$ as the closure of  the projection of the sets $\{ k\leq -b \}\cap S$ for 
$b$ large enough with $S$ defined in Section \ref{ss:escape}. 
We take for $V_0$ a neighbourhood of $K_-$ in $S^\star M$ which is
of the form $\{ {\rm exp}(u {\cal Y}_r )(m)|m \in V_1, |u|\leq a  \}$.
If $a$ is small enough the vector field  ${Y}_r $ is transversal and outgoing at the boundary of $V_0$.
Because ${Y}_r (h)= h \{ h, k \}$ and $\{ h, k \} \geq \delta >0$. 
Hence we get a repellor $L_-:=\cap_{t\leq 0} \psi_t (V_1)$. The repellor $L_-$ contains $K_-$ and being invariant by the dynamics
of ${\cal Y}_r$  restricted
to $Z_0$ is equal to $K_-$.
We then take for $U_1$ a small neighbourhood of $V_0$ in $\overline{T^\star M}$ 
 and we get item 2 by taking for $U$ in the definition of a radial source
the intersection  $U_0\cap U_1$.  $\square$
%%%%%%%%%%%%%%%%%%%%%%%%%%%%%%%%%%%%%%%%%%%%%%%%%%%%%%%%%%%%%%%%%%%%%%%%%%%%%%%%%%%%%%%%%%%%%%%%%%%%%%%%%%%%%%%%%%%%%% 

\section{Applying Mourre's theory}\label{sec:mourre}

Let us first recall some results of \cite{CSR-18}.
Let us fix $\omega =0$ for simplicity and assume that there exists an escape function $k$ on the energy
shell $\Sigma _0$. Then $k$ can be extended to $T^\star M\setminus 0$ as an escape function in the cone $|h|\leq a $ with some $a>0$.
Let $K$ be a self-adjoint operator of degree $1$ of principal symbol $k$. Using the `` G\r{a}rding's inequality'' (see \cite{Fo-89}
pp 129--136), one gets that
$K$ is a conjugate operator in the sense of Mourre: if $J$ is as small enough open interval containing $0$ and 
$\pi_J $ is the spectral projector of $H$ associated to the interval $J$, then
\[i \pi _J [H,  K] \pi_J \geq c \pi_J + R \]
where $c>0$ and $R$ compact.
Moreover the operator $H$ is $K$-smooth, i.e. the map
$t \ra e^{itK }H e^{-itK }$ is smooth with values into the  bounded self-adjoint operators.
Let us define the $K$-Sobolev spaces, denoted
${\cal H}_K^s$,  in the usual way using the $s$-powers of 
$(1+K^2)^{\ha }$. The usual Sobolev spaces will be denoted by ${\cal H}^s$. Let us give  a comparison between
 the $K-$Sobolev spaces and the usual ones. There is a shift in the exponents due
to the fact that the pseudo-differential calculus does not apply to non elliptic operators like $K$.  
\begin{lemm}  If $f\in {\cal H}^1$, then $f\in {\cal H}_K^s$ for any $s\leq 1$.
If $f\in {\cal H}^{-1}_K$, then $f\in {\cal H}^{-s}$ for any $s\leq -1$.
\end{lemm}
\begin{demo}
If $f\in {\cal H}^1$, $<(1+K^2)f|f> <\infty $ because $K^2$ is a pseudo-differential operator of order $2$ and hence $f\in {\cal H}_K^1$.
The other inclusion follows by duality w.r. to the $L^2$ product.

\end{demo}

It follows then from Mourre theory \cite{Mo-81,Mo-83,JMP-84,Ge-08} that
\begin{theo}[Mourre]
The operator $H$ has a finite number of eigenvalues in $J$, they have finite multiplicity.
Assuming that $0$  is not an eigenvalue,
the resolvent $(H-z)^{-1} $ defined for $\Im z>0 $ admits a boundary value
$\omega \ra (H-\omega -i0)^{-1}$  for $\omega $ real close to $0$ which, for any $\epsilon >0$, 
  is H\"older continuous for some positive H\"older exponent, depending on $\epsilon$, 
from the Sobolev space ${\cal H}_K^{\ha +\epsilon} $ into  ${\cal H}_K^{-\ha -\epsilon} $
for all $\epsilon >0$.

Moreover, if $\Pi _- $ is the spectral projector on the negative part of the spectrum of $K$, then,
if $f\in {\cal H}_K^{1 +\epsilon} $, then 
$\Pi_- \left( (H-i0)^{-1}f \right) \in L^2$. 
\end{theo}

It follows then in our context:
\begin{theo}[\cite{CSR-18}]\label{theo:CSR}
Assuming the existence of an escape function at $\omega =0$ and that $0$ is not an eigenvalue of $H$,
then the solution $u(t)$ of the forced wave equation (\ref{equ:forced}) with a smooth forcing $f$ can uniquely be written as 
\[ u(t)=u_\infty +\eta (t) +r(t) \]
where
\begin{itemize}
\item $u_\infty =(H-i0)^{-1}(f) $ belongs to ${\cal H}_K^{-\ha -\epsilon} \subset {\cal H}^{-1}$ for all $\epsilon > 0$

\item $\eta (t)\ra 0 $ in  ${\cal H}_K^{-\ha -\epsilon}\subset {\cal H}^{-1} $ for all $\epsilon > 0$
\item The function $t\ra r(t)$ is bounded in $L^2$ has a Fourier transform vanishing near $0$
\item $\| u(t) \|^2_{L^2}\sim c t $ as $t\ra +\infty $ with in general $c>0$.
\end{itemize}

\end{theo}

\section{Using radial source and sink propagation results}\label{sec:DZ}
\subsection{Wavefront set of $u_\infty $}\label{sec:WF}

We will now derive results on the distribution $u_\infty $ using  the  radial
propagation estimates of  Dyatlov-Zworski, based on earlier ideas of Richard Melrose \cite{Me-94}
and Andras Vasy \cite{Va-13},   and get 
\begin{theo}\label{theo:WF} 
The wavefront set of $u_\infty $ is contained in the cone $\Gamma_+$.
\end{theo} 
\begin{demo}
The result follows from the argument explained in the revised version of \cite{DZ-18}, section 3.1.
This use only the fact that $K_-$ is a source (see Section \ref{sec:source}). 
They introduce an operator 
$<D>$ which is elliptic  self-adjoint invertible of degree $1$. We choose it so that its  principal symbol near $\Gamma_-$ is $|k|$. 
They  introduce then 
\[ v_\epsilon :=  <D>^{-\ha} (H-i\epsilon)^{-1} <D>^{-\ha}( g) ~,  \]
with $g= <D>^{\ha}( f)$ and $u_\epsilon =(H-i\epsilon)^{-1}(f)=<D>^{\ha}v_\epsilon $.
Using a refined version of the Theorem E.54 of \cite{DZ-17},
they show  that there exists  $A$,  elliptic near $\Gamma _-$  of degree $0$,
 so that, for any $s$, the norms     $\| Av_\epsilon  \|_s$ are uniformly bounded in $\epsilon >0$. We need to use here, 
in the inequality (3.2) of \cite{DZ-18}, that  $ \| v_\epsilon  \|_{-N} $ is bounded; we know it from Mourre theory for
$N\geq 1$.
Passing to the limit which is known to exists in ${\cal H}^{-1}$ by Theorem \ref{theo:CSR}, we get
that $u_\infty $ is smooth near $\Gamma _-$. 
The usual propagation of singularities applied
to the equation $Hu_\infty \in C^\infty $ gives the result.  

\end{demo}

\begin{prop}
If $Hu=0$ and $u \in L^2(M)$, then $u$ is smooth.
\end{prop}
\begin{demo} It follows from Exercice 33 in Appendix E7 of \cite{DZ-17}, that $u$ is smooth near $\Gamma _-$
and changing  $H$ into $-H$, $u$  is also smooth near $\Gamma _+$. 
\end{demo}

\begin{rem} In the case $n=2$, not all closed conical invariants subsets of $\Gamma _+$ can be wavefront sets of some $u_\infty$.
If the wavefront set contains the line generated by a (ws)saddle point, it contains also one of the 2 branches of the
associated  unstable manifold and hence, being closed,  also an attractive invariant set. This is proved in the paper \cite{GS-77}
at least for generic cases.
\end{rem}

%%%%%%%%%%%%%%%%%%%%%%%%%%%%%%%%%%%%%%%%%%%%%%%%%%%%%%%%%%%%%%%%%%%%%%%%%%%%%%%%%%%%%%%%%%%%%%%%%%%%%%%%%%%%%%%%
\subsection{Sobolev regularity of $u_\infty $}

We saw in Section \ref{sec:mourre}   that $u_\infty $ belongs to ${\cal H}^{-1}$. Let us show that the radial sink
estimates of \cite {DZ-17} allows to get 
\begin{theo}Under the assumption of existence of an escape function, we have, for all $\epsilon >0$, 
$u_\infty \in {\cal H}^{-\ha -\epsilon }$.  
\end{theo}
\begin{demo}
We use the fact that $K_+$ is a sink as defined in \cite{DZ-17}, definition E.52.:
 this is proved exactly the same way that we proved that $K_-$ is a source 
in Section \ref{sec:source}, or just by reversing the orientations. 
We use then  Theorem E.56 of \cite {DZ-17} directly for the operator $H$ knowing already that $u_\infty $ is smooth away of $\Gamma _+$.
Replacing $<\xi >$ by $<k_+>$ we see that the threshold condition (E.5.44)  is satisfied for $s<-\ha $.  

\end{demo}

%%%%%%%%%%%%%%%%%%%%%%%%%%%%%%%%%%%%%%%%%%%%%%%%%%%%%%%%%%%%%%%%%%%%%%%%%%%%%%%%%%%%%%%%%%%%%%%%%%%%%%%%%%%%%%%%%%%%%% 

%%%%%%%%%%%%%%%%%%%%%%%%%%%%%%%%%%%%%%%%%%%%%%%%%%%%%%%%%%%%%%%%%%%%%%%%%%%%%%%%%%%%%%%%%%%%%%%%%%%%%%%%%%%%%%%%%%%%%% 

\section{The 2D case}\label{sec:2D}
In this Section $n=2$.
\subsection{Morse-Smale foliation}
\begin{defi}
A hyperbolic singular point of ${\cal F}$  is said {\rm weakly stable} if the trace
 of the linearization of any  smooth vector field generating ${\cal F}$ is $<0$. We define similarly {\rm weakly unstable }
hyperbolic singular  points. We denote these properties respectively (ws) and (wu).
\end{defi}
Note that if $dh \ne 0 $ on $\Sigma _0 $,  any  saddle points is  either weakly stable or weakly instable depending on 
the fact that ${\cal X}_h $ is pointing  to the infinity  or not, this follows from equation 
(\ref{equ:equ-div}) where $a\ne 0$. 

Let us recall that a vector field on a surface is {\it Morse-Smale} if the non wandering points are singular hyperbolic points
and closed hyperbolic cycles and there is no saddle connection,
i.e. there is no leave  whose both limit points are  saddle points.
 We extend this definition to oriented foliations of surfaces by choosing  any vector field generating the
foliation. 

\begin{theo}\label{theo:main}
Let  $n$ be equal to $2$.  Let us assume that the foliation ${\cal F}$
is Morse-Smale.
Then  there exists an escape function. The set 
$K_+$ is the union of all the attracting cycles and points 
and all  the unstable manifolds of the ws-saddle points.
The set  $K_-$ is  constructed in  a similar way. 
\end{theo}
\begin{rem}
Any generic foliation of a closed surface satisfies the previous properties:
Mauricio  Peixoto proved in the sixties   that  Morse-Smale vector fields on surfaces  are generic,
  see \cite{PdM-82}, Chapter 4,  for a detailed proof. 
As pointed out to me by Sylvain Courte, this genericity property extends to our context, i.e. to singular foliations of a 
surface embedded in a contact manifold,  as it is proved 
in the PhD thesis of Emmanuel Giroux \cite{Gi-91}, Lemme 1.3. 
\end{rem}

{\it Proof--.}
Note first that $K_+$ and $K_-$ are compact. They are also  disjoint because there is no saddle connection.

Let us  prove that  $K_+ $ is an attractor. Let $K_0$ be the union of the attracting component of $K_+$. The compact 
$K_0$ itself is an attractor. 
Let us assume for simplicity that there exists an unique (ws) saddle-point $b$.
Near $b$ the foliation has a local normal form: the level sets of the function $xy$ in a ball $B$ contained in $\R^2_{x,y}$
with the orientation given by  $x\pa_x -y\pa _y $. 
Let us consider a \nghbd $U_0$ of  $K_0$  satisfying the conclusion of Proposition 
\ref{prop:att}.
 The bassin of $K_0$ is the complement in $Z_0$ of
the  union of all unstable cycles and all the stable  manifolds of the saddle points.
  In  particular by taking $\phi_{-T} (U_0)$ with $T$ large enough
 instead of 
$U_0$ one can assume that $U_0$  contains $L:=\{ |x|\geq a, |y|\leq b \}\cap B $ with $a,b>0$.
Let us take now for the \nghbd of $K_+ $ the set
$U:=U_0 \cup L$. Clearly $\cap_{t\geq 0} \phi_t (U)=K_+$.

\begin{rem}
$K_0 \cup \{b \} $ is not an attractor! 
\end{rem}

Let us fix a density $d\mu $ on $Z_0$ and construct  a vector field $W$ generating ${\cal F}$  near $K_+$ whose 
divergence is non positive on $K_+$. First, we construct  a vector field $W_b$ with ${\rm div}(W_j)<0 $  in some neighbourhood
$U_b$ of  each (ws) saddle point $b$.
We construct also (see Appendix \ref{app:liap}) a vector field $W_a$ in the basin of each attractive cycle or point $a$ with non positive
divergence. 
Let us choose a  positive  function $l_a$ tending to $+\infty $ at the boundary of the basin of $a$.
Then, for $L_a$ large enough the set $\{ l_a\geq L_a  \}$ intersects the unstable manifolds $Y_j$ of each  (ws) saddle point $b_j$ inside
$U_{b_j}$. 
We choose $\chi_a \in C_o^\infty (\R,[0,1])$ so that $\chi_a (s)=1$ for $0\leq s \leq L_a$ and 
$\chi_a'(s) \leq 0 $ for $s\geq 0$.
Then we take,
\[ W= \sum _a \left( (\chi _a \circ l_a) W_a +C \sum _{b_j  (ws)}\psi_j W_{b_j} \right) \]
where $\psi_j $ satifies 
\begin{itemize}
\item $\psi_j \in C_0^\infty (U_{b_j}, R_+ )$
\item $\psi_j =1 $ on $\{ l_a \geq L_a \} \cap Y_j $
\item $d\psi_j (W) \leq 0 $ on $Y_j\cap U_{b_j}$
\end{itemize}
and 
 $C>>1$.
This smooth vector field is  well defined near $K_+$
 and has negative divergence on $K_+$.
$\square $
%%%%%%%%%%%%%%%%%%%%%%%%%%%%%%%%%%%%%%%%%%%%%%%%%%%%%%%%%%%%%%%%%%%%%%%%%%%%%%%%%%%%%%%%%%%%%%%%%%%%%%%%

\subsection{Lagrangian distributions associated to hyperbolic closed leaves}

Let $\Gamma \subset T^\star X \setminus 0$  be a conic component of $\Gamma_+$  generated by a closed hyperbolic cycle $K_{+,0}$
  of the foliation ${\cal F}$.
The cone  $\Gamma $ is a conic Lagrangian submanifold of $T^\star X\setminus 0$: the Euler identity implies
$\omega ({\cal X}_h , p \pa _p )=0$.
A theorem of Alan Weinstein \cite{We-71} implies that there is an homogeneous
canonical transformation $\chi $ defined in a conic neighbourhood $C$  of
$\Gamma $ whose image is a conic neighbourhood of the
zero section of $T^\star \Gamma $ and so that $ \chi (\Gamma ) $ is
the zero section of  $T^\star \Gamma $. More precisely $\chi $ restricted to $\Gamma $ identifies $\Gamma $ to the zero section of its own
cotangent bundle. 
Taking polar coordinates $(x,\eta )\in (\R/\Z) \times (\R_+\setminus 0) $ on the cone $\Gamma $,
the cotangent bundle of $\Gamma $ admits coordinates
$(x,\eta; \xi, y )$ with the symplectic form $d\xi \wedge dx + dy\wedge d\eta $. Note that they are not the symplectic cocodinates
of $T^\star X$, but those of $T^\star \Gamma $! Let $X_0$ be defined as $X_0:= ( \R / 2\pi \Z)_x \times \R_y $.
The symplectic map $(x,\eta;\xi, y) \ra (x,-y; \xi,\eta) $ from
$T^\star \Gamma $ onto $T^\star X_0$ identifies $T^\star \Gamma $ with $T^\star X_0 $.
With this identification, $\Gamma $ is moved into  $\Gamma _0=\{ y=0,\xi =0 \}$ which is the conormal bundle of the circle 
of $\gamma _0 \subset  X_0$ defined by $y=0$. 
The  Hamiltonian vector field ${\cal X}_0  $ of $h_0:=h\circ \chi^{-1}$
preserves $\Gamma_0  $.
Along $\Gamma _0 $, it is then given
by ${\cal X}_0 =\pa _\xi h_0  \pa _x -\pa _y h_0  \pa _\eta  $
and there $\pa _x h_0=\pa _\eta h_0=0$.  Because the foliation ${\cal F}$ is non singular near  $K_{+,0}$,
we have 
$ \pa _\xi h_0\ne 0 $. 
 Hence the image of the energy shell $\Sigma _0 $ is given
by $\xi/\eta = F(x,y) $. The projection $\pi : Z_0 \rightarrow X_0 $ is a local diffeomorphism   near $K_{+,0}$. 
Because it is a diffeomorphism  on the cycle $K_{+,0}$, it is even a global diffeomorphism.

Using the tools introduced by Alan Weinstein in \cite{We-75}, we can build
 a FIO microlocally unitary $U:L^2(X) \rightarrow L^2
 (X_0,M)$
with $M$ a flat bundle, called the Maslov bundle,  so that
$ UH U^\star -K $ is smoothing in $C$ and 
$\sigma _p (K)= h\circ \chi^{-1}$,
${\rm sub }(K)=0$.
We are then reduced to the  case already studied in \cite{CSR-18} where the projection of $\gamma $  onto $M$ is a diffeomorphism. 

This proves, following then  \cite{CSR-18}, 
the 
\begin{theo} If $\Gamma $ is a component of $\Gamma _+$ generated by a closed hyperbolic stable cycle
of ${\cal F}$, 
 the distribution $u_\infty $ is microlocally near $\Gamma $ a Lagrangian distribution.
\end{theo}

%=================================================================================================================

\section{The 3D case with $S^1$ invariance}\label{sec:3D}

Quite often in physical situations, there is an invariance of the problem by rotation or translation: internal waves in some canal
 \cite{ML-95}, inertial waves inside the earth or some stars \cite{RV-18}, \dots
We will study the case where $M=N_q\times S_\theta^1$ is a 3-manifold with the canonical action of
   $S^1$ by translation on the second factor.
We denote by $(q,p;\theta ,\tau )$ some local canonical coordinates on $T^\star M$ and 
  assume that $N$ is equipped with a smooth density $|dq|$
and $M$ with $|dq d\theta |$. 
Let us give a smooth  Hamiltonian $h=h(q,p,\tau  )$, homogeneous of degree $0$,  on $T^\star M \setminus 0$
and a   self-adjoint pseudo-differential operator of degree $0$,
 $H$, of principal symbol $h$,  acting  on $L^2(M,|dq d\theta |)$.
We assume that $H$
 commutes with the $S^1$-action. 
The operator $H$ is then a direct sum  of operators on $M$:
\[ H=\oplus _{n\in \Z} H_n \]
where 
$H_n$ acts on $L^2 (N,|dq|)$ as a self-adjoint pseudo-differential operator of principal symbol
$h_n(q,p):=h(q,p,n )$ which is also equal to $h(q,p/n,1)$ if $n \ne 0$.

The spectrum of $H$ is clearly the closure of the union of the spectra of the $H_n$'s. 

\subsubsection{Spectra of $H$ and the $H_n$'s}

Let us define
$h_0 (q,p):=h(q,p,0)$
and $h_1 (q,p)=h(q,p,1)$. Note that $h_1$ is a smooth symbol of degree $0$ on $T^\star N$  which is asymptotic to $h_0$ at infinity. 
The essential spectrum of $H$ is the interval 
$I_\infty:=[a_\infty , b_\infty ]$ where $a_\infty =\inf h_1 $
and $b_\infty =\sup h_1 $.
The essential spectrum  of the $H_n $'s is quite different: from the identities
\[ h(q,p, n)=h\left(q,\frac{p}{|p|},\frac{n}{|p|}\right)=h_0(q,p)+O\left( \frac{1}{|p|} \right)~,\]
 one gets that the principal symbol of $H_n $ is $h_0$. 
Hence the essential spectrum   of any of the  $H_n $'s  is    $I_0:=[a_0, b_0]$ where 
$a_0 =\inf h_0$ and $b_0 =\sup h_0$.
Note that we have
$I_0 \subset I_\infty $ and they are often identical in the applications to physical problems. 

We are interested in more precise properties of the spectra:
we claim that, in $I_\infty \setminus I_0$,
 the spectrum of $H$ is   pure point dense, i.e. there is a basis of $L^2$ pairwise orthogonal
eigenfunctions. Moreover the eigenvalues  of $H_n$   obey a Weyl rule when $n \ra \infty $. 
One expects  that the spectrum has no embedded eigenvalues in the interior of  $I_0$. But quasi-modes
of the type ``well in an island'' are possible if the dynamics of $h_1$ has stable bounded invariant sets
(see Section \ref{sec:class}).

\begin{theo}[Weyl law]\label{theo:weyl}
 The spectra  $\sigma (H_n )$ of the operators $H_n$ in $I_\infty \setminus I_0$ are discrete. 
 For any compact interval $J$ included in $I_\infty \setminus I_0 $,
we have
\[ \# \{ \sigma (H_n ) \cap J \} \sim _{n \ra \infty } 
\frac{n^2}{4\pi ^2} {\rm vol}\left( \{ q,p|h_1(q,p)\in J \} \right) 
\]
where the volume is defined with the Liouville measure on $T^\star N$
and the eigenvalues of $H_n $  in $J$ are counted with multiplicities. 
\end{theo}
{\it Proof.--}
The full symbol of $H$
writes
\[ \tilde{h}= h(q,p,\tau ) +\sum _{j=1}^\infty k_j (q,p,\tau ) \]
with $k_j$ homogeneous of degree $j$.
Hence $H_n$ can be viewed as a semi-classical pseudo-differential operator on $N$ of semi-classical symbol
\[ \tilde{h}_n= h_1(q,\hbar p ) +\sum _{j=1}^\infty \hbar^j k_j (q,\hbar p,1 ) \]
with $\hbar =1/n$. 
The Theorem follows hence from the semi-classical Weyl
asymptotics. $\square $
\subsubsection{Classical dynamics}\label{sec:class} 
We will assume that the frequency $\omega =0$ is fixed and the 2D Hamiltonian $h_0(q,p):=h(q,p,0)$ admits an escape function.
We will look at the dynamics of $h_1:=h(q,p,1)$.
Note that the dynamics of $h$ reduces on each set $\tau =a$ with $a\ne 0$ to that of $h_1$ by some simple rescaling 
of the time.
Moreover
\[ \lim _{p\ra \infty } h_1(q,p)= h_0(q,p) \]
Near infinity the dynamics stil admits an escape (Liapounov function) and hence the orbits, if they come close enough to
infinity, will converge to $K_+$ at $+\infty $ and $K_-$ at $-\infty $.
The dynamics $t \ra \phi_t $  of $h_1$ is hence complete.
We split the phase space into 3 pieces:
$T^\star M = \Omega \cup C_+ \cup C_-$
where
\begin{itemize}
\item  $\Omega $ is the set of $(q,p)$ so that 
$\phi_t (q,p )\ra K_\pm $ as $t\ra \pm \infty $
\item $C_+$ is the set of  $(q,p)$ so that $\phi _t (q,p)$ stays bounded for $t\geq 0 $
\item $C_-$ is the set of  $(q,p)$ so that $\phi _t (q,p)$ stays bounded for $t\leq 0 $
\end{itemize}
Finally, we define $C:=C_+ \cap C_- $ the set $(q,p)$ so that $\phi _t (q,p)$ stays bounded for $t\in \R $.
In the literature, $C$ is called the {\it trapped set}.

It could happen that $C$ supports some quasi-modes associated to the semi-classical parameter $1/n $. 
Generically, these quasi-modes are not close to true $L^2-$eigenfunctions because such eigenfunctions  do  not exist. 
There are stil visible in the wave dynamics for a very long time...

\section{Open problems}\label{sec:pb}

%%%%%%%%%%%%%%%%%%%%%%%%%%%%%%%%%%%%%%%%%%%%%%%%%%%%%%%%%%%%%%%%%%%%%%%%%%%%%%%%%%%%%%%%%%%%%%%%%%%%%%%%
There are stil many open problems. Let us describe a few of them:
\begin{itemize}
\item How does the spectral picture extends outside the  intervals with a.c. spectra?
This problem is already not solved in the simple case where $Z_0$ is a 2-torus, assuming the existence of  a global transversal
to the foliation, and the Poincar\'e map loses its hyperbolicity in a generic way. 
\item More generally, can we study what happens at the critical values of $h$ assuming  that this
function is Morse or even Morse-Bott on $S^\star M$?
\item What can we do in the case of a manifold with boundary? In particular, can we say something in the case of a polygon
which is studied in the experiments of the Thierry Dauxois's team \cite{Br-16}.
\item Prove the generic absence of embedded eigenvalues.
\item Consider the {\it viscous case}, namely the forced equation
\begin{equation}\label{equ:viscous}
\frac{du}{dt}+ iHu-  {\sigma}\Delta u   =fe^{-i\omega t},~u(0)=0~. \end{equation}
where $\sigma $ is a positive number and  $\Delta $ is the  Laplacian associated to some Riemannian metric on $M$. 
Study the  ``small viscosity''  limit $\sigma \ra 0$? In particular, do the limits 
$\sigma \ra 0^+$ and $t\ra + \infty $ commute?
\item There is a discrete analogue of Mourre's theory for unitary maps, see for example 
\cite{FRT-13}.  What can be said from the spectral theory of the 
unitary action of a diffeomorphism of a closed manifold on half-densities?
For example, what is  the spectral theory of a diffeomorphism of the circle with irrational rotation number which is  not
$C^1$-conjugated to a rotation? 

\end{itemize}

%%%%%%%%%%%%%%%%%%%%%%%%%%%%%%%%%%%%%%%%%%%%%%%%%%%%%%%%%%%%%%%%%%%%%%%%%%%%%%%%%%%%%%%%%%%%%%%%%%%%%%%%%%%

\section*{Appendices}
\appendix

\section{ Divergences}
\subsection{Formulae}\label{app:form}

Let us give a smooth vector field $W$ whose flow is denoted $\phi_t,~t\in \R $  and a smooth density $d\mu $.
The divergence of $W$ with respect to $d\mu$ is the function 
defined by
\[ {\rm div}_{d \mu }(W):= \frac{{\cal L}_W d\mu }{d\mu }\]
where the Lie derivative 
${\cal L}_W d\mu $ is defined
by ${\cal L}_W d\mu :=\frac{d}{dt}_{|t=0} \phi_t ^\star (d\mu )$.
Cartan's formula gives
\[ {\rm div}_{d \mu} (W)=\frac{ d(\iota (W)d\mu ) }{d\mu }\]
where $\iota(.)$ is the inner product. 
In particular, we get the usefull formulae
\[ {\rm div}_{d\mu }(fW)=df(W) + f {\rm div}_{ d\mu }(W)\]
\[ {\rm div}_{gd\mu  }(W)=\frac{dg(W)}{g} +  {\rm div}_{d \mu} (W)\]

%%%%%%%%%%%%%%%%%%%%%%%%%%%%%%%%%%%%%%%%%%
\subsection{ Extending vector fields with negative divergence}\label{app:liap}

\begin{lemm}
Let us assume that the invariant compact $K$ admits a smooth (Liapounov) function $l$ defined in the basin $B$  of $K$
with $dl(W) <0 $ outside $K$ and  $l(K)=0$ and $l \ra +\infty $ at the boundary of $B$
(this is the case in particular if the attractor $K$ is hyperbolic).
If the vector field $W$ satisfies ${\rm div}_{d\mu } (W) <0 $ in some open neighbourhood $V$  of $K$, then
 there exists a vector field $W_1=F W$ in  $B$,  so that
$F>0$ and ${\rm div}_{d\mu } (W_1) <0 $ in $B$. \end{lemm}
{\it Proof--.}
 Let us choose $r>0$ so that  $\{ l\leq r \} \subset V$.
It is enough to take $F=1$ in $\{ l\leq r \}$ and, for any $x\in B$ with $l(x)=r$   and any  $t\geq 0$,   
\[ F(\phi_t (x)) := e^{ \int _{0}^t \Phi  (\phi_s (x)) ds }\]
with $\Phi $ smooth, $\Phi =0$ near $l(y)\leq r$  and, for all $y$ with $l(y)>r$,  $\Phi (y)< - {\rm div}_{d\mu } (W)(y)$.  $\square$

\section{Attractors and their basins}\label{sec:dyn}:

We give here some useful definitions and elementary properties of dynamical systems.
We consider a smooth closed manifold $X$ with a smooth  vector field 
$V$ whose flow is the 1-parameter group of diffeomeorphisms of $X$ denoted by $\phi_t,~t\in \R$.
The definitions and statements are taken from the reference \cite{Hu-82}.
We have the following
\begin{defi}
\begin{enumerate}
\item 
If $K\subset X$ is a compact  invariant set, i.e. a subset of $X$ preserved by the flow,
$K$ is called an {\rm attractor} if there exists an open neighbourhood  $U$  of $K$ in $X$ so that
$K=\cap _{t\geq 0} \phi_t(U)$.
 \item 
If $K$ is an attractor, the {\rm basin} of $K$ is
the set of  points $x$ so that $\phi_t(x) \ra K $ as $t\ra +\infty $. 
\item A point $x\in X$ is wandering if there exists a \nghbd ~$U$ of $x$ so that
$\phi_t(U) \cap U=\emptyset $ for $t$ large enough. 
\end{enumerate}
\end{defi}

The set of wandering points is open.
The basins are open subsets of $X$. 
We will need the following properties (Lemma 1.6 of \cite{Hu-82}):
\begin{prop}\label{prop:att} 
If $K$ is an attractor, and $V$ a neighbourhood of $K$, there exists an open set $U$
satisfying
\begin{enumerate}
\item $K\subset U \subset V$
\item $\cap _{t\geq 0}\phi_t (\bar{U})=K$
\item For all $t\geq 0$, $\phi_t(U)\subset U$
\end{enumerate}
The convergence of $\phi_t(m)$ to $K$ is uniform on every compact subset of the basin of $K$.
\end{prop}

The previous sets  are the same for $V$ and $fV$ where 
$f:X\ra ]0,+\infty [ $ is smooth. They can therefore be defined for a 1D oriented foliation generated by a smooth vector field.  
In particular the open set $U$ of the previous proposition is independent of $f$.

\bibliographystyle{plain}

\end{document}